# Perception proximale et distale à l'aide du dispositif de Lenay

*S. Hanneton, D. Taleb Kachour, M.M. Ramanantsoa, B. Hardy et A. Roby-Brami*

## Résumé

Nous présentons dans cet article des résultats expérimentaux obtenus lors de l'utilisation du dispositif de Lenay pour la localisation (perception distale) de cibles et la perception (proximale) de l'orientation d'un cylindre dans un plan. Cette seconde expérience emploie une version "virtuelle" du dispositif de Lenay. Ces résultats sont ici présentés au titre d'illustrations, l'objectif principal de ce travail étant de proposer des orientations méthodologiques et théoriques pour l'étude des processus sensori-moteurs et cognitifs impliqués dans la perception.

## Introduction

### *Le dispositif de Lenay*

Le dispositif de Lenay, dans sa forme initiale, est un dispositif de suppléance perceptive élémentaire. Il est constitué d'un capteur unique couplé par un dispositif électronique à un unique stimulateur fonctionnant en tout ou rien. Un des prototype proposé [LENAY 1997] comporte comme capteur un phototransistor couplé à un vibrotacteur entrant en action lorsque l'intensité lumineuse parvenant au phototransistor dépasse un seuil ajustable. Plusieurs dispositifs peuvent être assemblés et fonctionner en parallèle par l'association de plusieurs capteurs identiques possédant des champs récepteurs différents couplés à des stimulateurs identiques activant des récepteurs sensoriels différents. Cette association de dispositifs identiques permet l'étude de l'influence du parallélisme sur les capacité perceptives des sujets portant de tels dispositifs [SRIBUN 2002]. Le dispositif de Lenay peut être adapté pour permettre une perception proximale ou bien une perception distale des objets. Avec un dispositif de suppléance perceptive proximale, les capteurs déclenchent les stimulations sensorielles lorsqu'ils rentrent en contact avec l'objet (le champ récepteur est centré sur le capteur et restreint). Dans le cas d'un dispositif de suppléance perceptive distale, les capteurs activent les stimulateurs lorsque l'objet entre dans leur champ récepteur.

Différentes formes du dispositif de Lenay ont été utilisées pour étudier les capacité de sujets à discriminer ou reconnaîre des formes ou caractères [HANN 1998], à reproduire l'orientation de segments de lignes [LENAY 2002], , à localiser des cibles [HARDY 2000].

### *Dispositif de Lenay et suppléance sensorielle*

Ce dispositif introduit un couplage entre des récepteurs sensoriels artificiels (les capteurs) et des stimulateurs produisant une activation de récepteurs sensoriels biologiques. Le système de couplage électronique permet donc de réaliser la transduction d'une forme d'énergie en une autre (photonique en mécanique par exemple). Un tel dispositif détourne donc le mécanisme de transduction réalisé par les récepteurs biologiques, permettant ainsi par exemple d'induire des sensations tactiles liées à des variations de l'énergie photonique, ou des sensations auditives liées à des variations de l'énergie thermique. Chez un sujet "normal", c'est-à-dire possédant des systèmes sensoriels fonctionnels, le dispositif de Lenay peut ainsi donner accès, via l'apprentissage, à de nouveaux modes de perception. Chez un sujet doté d'une déficience sensorielle, ce dispositif permet de suppléer le sens déficient, en utilisant un système sensoriel fonctionnel pour réaliser la transduction inopérante (voir figure 1).

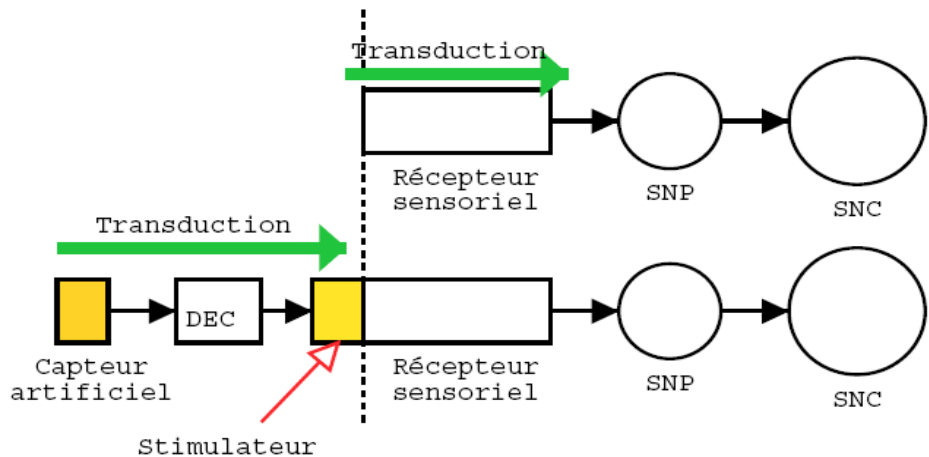

*Figure 1: Les dispositifs de suppléance sensorielle peuvent être considérés comme réalisant une transduction artificielle. Cette transduction réalise la transformation et l'amplification d'un type d'énergie (par exemple photonique) en un autre (par exemple mécanique). Abbréviations : SNP = système nerveux périphérique, SNC = système nerveux central, DEC = dispositif électronique de couplage. La ligne en pointillés figure la limite corporelle.*

C'est en ce sens que le dispositif de Lenay est un dispositif de substitution sensorielle ou plus précisément de suppléance perceptive. Le terme de substitution sensorielle, bien que génériquement utilisé pour des raisons historiques [BYR 1972], n'est pas adapté car ces dispositifs ne substituent pas véritablement un sens à un autre. Nous préférerons donc utiliser le terme de suppléance perceptive. Le dispositif de Lenay n'est pas une réalisation technique en soi bien que plusieurs prototypes furent développés. Il s'agit plutôt d'un paradigme qui défini une "brique de base" permettant d'unifier la description des multiples systèmes de suppléance perceptive existant. Ainsi, le TVSS de Paul Bach Y Rita est composé de l'association en parallèle de nombreux dispositifs réalisant la transduction d'une énergie photonique en énergie mécanique [BYR 1970].

A la condition d'étendre le dispositif de Lenay au pilotage d'une stimulation sensorielle continue (et non en tout ou rien), on peut décrire le système TheVoice de P. Meijer [MEIJER 1992] comme étant constitué de l'association de multiples dispositifs réalisant la transduction d'une énergie photonique en énergie sonore. Le dispositif de Lenay est donc un outil théorique permettant d'aborder l'étude de la suppléance perceptive pas uniquement dans l'objectif d'optimiser des développements techniques, mais surtout du point de vue de la recherche fondamentale. Dans les paragraphes suivants, nous ne nous intéresserons qu'à la forme canonique du dispositif de Lenay, c'est à dire limitée à un capteur et un stimulateur (tactile en l'occurence).

### *Intérêt du dispositif pour l'étude de la perception*

Lorsqu'un sujet voyant aveuglé explore son environnement avec le dispositif de Lenay, il peut utiliser deux types différents de sensations. Le premier type de sensation, médié par le dispositif (vibreur tactile), est lié à la structuration de l'environnement lui-même. Les informations données par ce premier type de sensation sont extrêmement pauvres car réduite à chaque instant à un signal binaire (stimulation ou pas de stimulation). Cette sensation peut être d'ailleurs également a priori considérée comme un simple signal de renforcement. Le second type de sensations disponible est lié aux mouvements effectués par le sujet lors de l'exploration de l'environnement avec le dispositif. Il s'agit de sensations issues en particulier de la proprioception musculaire ou articulaire, du toucher, de l'appareil vestibulaire. Ces sensations sont supposées permettre au sujet de connaître la position de son corps dans l'espace, sa posture et les mouvements effectués[1]. Les informations conduites par ce second type de sensations sont donc extrêmement riches comparées à la pauvreté des informations fournies sur l'environnement lui-même. La capacité du sujet à percevoir la localisation, la forme d'un objet dépend donc essentiellement de son aptitude à combiner le savoir extéroceptif très limité fourni par le dispositif à la connaissance qu'il peut avoir du mouvement d'exploration qu'il est en train d'effectuer. L'étude de l'apprentissage et de l'utilisation du dispositif donne ainsi un éclairage tout à fait intéressant sur le rôle du mouvement et du sens du mouvement dans la perception.

L'importance de la contribution de l'action et en particulier des mouvements segmentaires dans la perception est à l'heure actuel tout à fait reconnu. Mais la plupart des auteurs attribuent au mouvement (implicitement ou explicitement) un rôle dans la reconstruction d'une hypothétique scène interne, supportée par une représentation interne éventuellement multimodale de l'environnement. Ce modèle suppose que le système nerveux central du sujet reconstruise dans un espace interne associé à un référentiel particulier (ce référentiel diffère selon les auteurs) une description de l'environnement d'après les stimulations sensorielles reçues lors des mouvements d'exploration successivement effectués.

Ce modèle de la scène interne est actuellement très discuté (voir par exemple [OREGAN 2001]) et d'autres hypothèses voient le jour. Nous n'entrerons pas dans ce débat mais nous nous servons ici avec beaucoup de précautions de ce modèle simplement dans l'intention d'éclairer l'intérêt du dispositif. Supposons que l'on demande à un sujet de reconnaître une forme simple en deux dimensions placée sur un plan qu'il explore avec un dispositif de Lenay doté d'un unique capteur et d'un unique stimulateur. La « couleur » $C(x,y)$ de chaque point du plan est fonction, du point de vue d'un observateur extérieur, des coordonnées x et y du capteur manipulé par le sujet. Pour reconstruire la forme de l'objet, le sujet effectue à chaque instant des mouvements qui peuvent être décrit, toujours du point de vue de l'observateur extérieur, par un vecteur vitesse $V(t)=[\frac{dx}{dt},\frac{dy}{dt}]$. A chaque instant t, le dispositif envoie une stimulation sensorielle $s(t)$ qui dépend de la position du capteur sur le plan. Supposons l'existence d'une scène interne homomorphe $\tilde{C}(\tilde{x},\tilde{y})$ utilisant également un référentiel cartésien bi-dimensionnel. Pour reconstruire l'objet exploré dans cette scène, le sujet n'a accès qu'à une estimation de la position $\tilde{M}(\tilde{x},\tilde{y})$ ou de la vitesse $\tilde{V}(\tilde{v}_x,\tilde{v}_y)$ du capteur dans le plan. Cette estimation est construite à partir du savoir proprioceptif que le sujet possède des mouvements exploratoires effectués. Mais le sujet n'a jamais accès à l'intégralité de la scène ni même à une large vue locale. La scène interne doit être reconstruite par un processus d'intégration temporel qui peut par exemple prendre la forme suivante :

$$\tilde{C}(\tilde{x},\tilde{y})=\tilde{C}(\int_0^t \tilde{v}_x(t)dt, \int_0^t \tilde{v}_y(t)dt)=s(t)$$

Cette équation montre que la capacité du sujet à reconstruire la forme de l'objet dans son hypothétique scène interne dépend crucialement de la précision de la connaissance qu'il a des déplacements et déformations de son corps dans l'espace lors des mouvements exploratoires. Si cette connaissance est imprécise ou si le processus d'intégration temporel est défectueux, alors la forme perçue de l'objet sera très

---

1. Il faut noter que le système nerveux central du sujet peut éventuellement utiliser des copies d'efférences des commandes motrices pour renforcer la connaissance qu'il a de l'action motrice en cours d'exécution.

dégradée ou même instable[2].

Une telle expérience a été conduite avec un stimulateur sensorielle fonctionnant en tout ou rien lors de tâches de reconnaissance de formes simples (carré, rond, triangles) ou de lettres [HANN 1998]. Le taux de réponses corrects de sujets voyants aveuglés s'est avéré supérieur à celui du hasard dans la plupart des cas. Ils sont donc capables de réussir en partie cette tâche. Mais c'est l'observation du comportement des sujets qui s'est avérée très informative sur les caractéristiques du processus d'intégration spatio-temporelle en jeu. Cette tâche est cognitivement lourde pour les sujets. Ils ont tous été conscient de la nécessité d'une très forte concentration pour accomplir chaque essai. D'autre part, lors de nombreux essais, les sujets se sont « perdus » dans l'image. Ils n'arrivent pas dans ce cas à reprendre contact avec la forme explorée. Il faut noter que cette perte de contact n'est pas temporaire, due seulement à un écart malheureux dans l'exploration de la forme, mais que les sujets ne savent réellement pas où aller (en haut, en bas, à droite, à gauche) pour retrouver le contact avec la forme[3].

Ce résultat est étonnant car l'on sait que la perception proprioceptive de la position ou du mouvement de l'extrémité du bras est relativement bonne dans les tâches de pointage ou de saisie [DESMU 1998]. Donc le sujet ne semble pas être lors de cette tâche dans un mode de fonctionnement lui donnant accès à la mémoire des positions de son bras dans l'espace de travail. Si ce n'était pas le cas, il lui suffirait simplement, lorsqu'il s'écarte trop de la forme, de revenir à une position précédente dans laquelle il a reçu une stimulation.

Enfin, l'observation des trajectoires exploratoires produites par les sujet s'est avérée également très informative. L'application de la formule précédente suppose qu'un simple balayage régulier du plan permet d'avoir accès à la forme. Or très peu de sujets utilisent cette stratégie. On voit au contraire se développer essai après essai une stratégie générique impliquant un suivi des contours de la forme par micro-balayage. Les sujets effectuent des allers-retours sucessifs en dehors et à l'intérieur de la forme selon un chemin suivant approximativement le gradient de sensation dans le plan. L'exécution de ces micro-balayages ne semble impliquer aucune charge cognitive. L'utilisation du dispositif de Lenay apporte donc un éclairage particulier sur les processus d'intégration spatio-temporels en cause dans la perception[4].

Cette expérience met en évidence l'implication de deux mécanismes de natures différentes : un mécanisme d'exploration locale peu coûteux en terme de concentration et relativement autonome (les micro-balayages) et un mécanisme impliquant une charge cognitive élevée impliqué certainement dans le liage et la reconstruction d'un percept cohérent. Ces deux niveaux, l'un de nature plutôt sensori-motrice et l'autre cognitive, fonctionnent de concert pendant la tâche mais toute disruption de leur coordination semble provoquer une telle perturbation que le sujet semble bloqué, incapable de reprendre de façon cohérente son activité exploratoire. Le dispositif de Lenay est donc un outil de choix pour étudier l'articulation entre sensori-motricité et cognition dans des tâches perceptives. Nous proposons dans les paragraphes suivant de présenter sommairement deux exemples de travaux mobilisant des outils issus de la psychologie expérimentale pour aborder cette question. Nous souhaitons décrire ces travaux dans l'objectif de soumettre quelques pistes méthodologiques dont l'application au études utilisant le dispositif de Lenay nous semble

---

2. Le sujet peut avoir l'impression subjective d'être au même endroit dans le plan et de recevoir une stimulation sensorielle différente de celle reçue lors de la première visite de cet endroit.

3. La sensation subjective ressentie par les sujets lors de cette perte est proche de celle que l'on peut avoir lorsque l'on explore une scène avec des jumelles. Au moment ou l'on met les jumelles, et bien que l'on connaisse l'environnement exploré, il est nécessaire de trouver un point de repère connu pour commencer l'exploration. Il n'est souvent pas facile de retrouver ce point de repère et il arrive parfois que l'on se perde dans l'image et que l'on soit obligé de revenir fréquemment chercher de tels points de repère par la suite.

4. Il faut noter ici que les résultats décrits ne concernent que la perception assistée par un dispositif technique nécessitant un apprentissage de son fonctionnement.

intéressante.

# Localisation/Discrimination de cibles avec le dispositif

### *Objectifs de cette expérience*

Le premier objectif de cette expérience est de quantifier les capacités de sujets voyants aveuglés à utiliser le dispositif de Lenay pour localiser des cibles tridimensionnelles placées sur un plan de travail. Nous avons reproduit quasiment à l'identique le protocole d'une expérience destinée à étudier les mécanismes cognitifs en jeu dans le pointage de cibles présentées visuellement ou proprioceptivement [ROSS 1995]. Le second objectif de cette expérience est en effet de déterminer dans quel type de référentiel les sujets codent la position des cibles. Cet hypothétique référentiel peut être en effet, selon la littérature, ego- ou exocentré, de nature proprioceptive ou plutôt visuelle.

La littérature montre que l'observation de la forme et de la disposition des ellipses de confiance décrivant la distribution des zones pointées par des sujets peut être très informative de ce point de vue. Nous ne présenterons ici qu'une partie des résultats expérimentaux (préliminaires) associés à cette étude. Enfin, nous comparons également les performances de plusieurs groupes de sujets dont une partie seulement est confronté à une phase préliminaire d'« apprentissage » actif de la prothèse que nous nommerons entraînement. Les questions auxquelles nous essyons de répondre sont donc les suivantes

- Les sujets sont-ils capables de localiser ou de discriminer des cibles avec ce dispositif et avec quelle précision ?
- L'introduction d'une période préalable d'entraînement avec la prothèse a-t-elle une influence sur cette précision ?
- Quelles sont les caractéristiques du référentiel hypothétique dans lequel la position de ces cibles est « codée » ?

### *Méthodes*

La prothèse sensorielle en elle-même est constituée d'un phototransistor placé au fond d'un tube noir lui donnant un angle d'ouverture d'environ 20 degrés couplé à un stimulateur tactile (électromagnétique) placé contre la face palmaire de la seconde phalange de l'index de la main droite. Le tube contenant le phototransistor est fixé par du sparadrap sur l'ongle de l'index de la main gauche. Le sujet explore donc la scène de la main gauche et reçoit des stimulations tactiles sur la main droite. Cette stimulation est d'amplitude constante et de 250 Hertz de fréquence. Elle est déclenchée lorsque l'intensité lumineuse captée par le phototransistor dépasse un certain seuil. Le sujet est assis sur un tabouret, le dos contre le mur devant un plan de travail éclairé depuis le plafond par un projecteur infrarouge. On place devant lui une cible constitué d'une demi balle de tennis recouverte de papier adhésif réfléchissant la lumière infrarouge.

L'expérience consiste à présenter au sujet la cible placée sur la table. Cette cible peut occuper 6 positions différentes (voir figure 2) sur la table diposée selon le groupe de sujets concerné soit en arc de cercle (groupe DIRECTION) ou alignée selon une droite fuyant le sujet (groupe DISTANCE). On lui demande d'explorer la scène avec la prothèse (main gauche) en conservant le contact de son dos avec le mur. Lorsque sa phase d'exploration est terminée, il ramène la main gauche au point de départ. Il doit ensuite pointer sur la table avec l'index droit à l'endroit estimé de la postion de la cible.

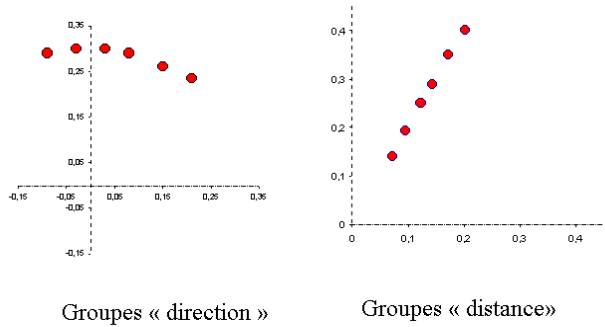

*Figure 2: Les cibles peuvent présenter deux types de localisation : en arc de cercle (expérience « DIRECTION ») ou alignées sur une droite fuyant le sujet (expérience « DISTANCE »). Des groupes différents de sujets passent les expériences.*

Ce pointage doit être accompagné de la prononciation du mot « Go ! » (condition GO) ou de la prononciation du chiffre (de 1 à 6) correspondant au numéro de la position estimée de la cible (condition CHIFFRE). Chaque sujet effectue 60 essais randomisés séparés en 3 phases de 20 essais (10 en condition GO et 10 en condition CHIFFRE). Au début de chaque phase de vingt essais, il est demandé au sujet d'apprendre proprioceptivement les 6 positions de cibles possibles. Pour cela, l'index de la main droite du sujet est placé par l'expérimentateur sur une des 6 positions possibles. L'expérimentateur prononce alors le numéro associé à la position puis ramène la main du sujet dans sa position initiale. Le sujet doit ensuite repointer avec l'index la position mémorisée. Puis une autre position de cible est choisie au hasard. Cette procédure est répétée jusqu'à ce que le sujet soit successivement capable de produire 10 résultats corrects (l'index pointe à moins de un centimètre de la position). Cette courte phase d'apprentissage proprioceptif a été introduite afin de s'assurer que le sujet possède une bonne cartographie proprioceptive des emplacements possibles des cibles et qu'il est capable d'associer correctement un numéro à chaque lieu.

L'expérience a concerné 5 groupes de 10 sujets (voir table ) Un groupe de sujets témoins aveugles de naissance a passé l'expérience DIRECTION. Nous avons proposé ensuite l'expérience à 4 groupes de sujets normaux aveuglés. Les deux premiers groupes de 10 sujets passent respectivement l'expérience DIRECTION et l'expérience DISTANCE comme elles sont décrites ci-dessus. Les deux derniers groupes suivent exactement le même protocole à une différence près. Nous leur avons en effet proposé une phase d'entraînement avant qu'ils ne commencent véritablement l'expérience. Durant cette phase d'entraînement le sujet est autorisé à manipuler seulement dix fois de suite lui-même la cible. Il place avec sa main droite la cible où il le souhaite sur la table puis place cette main sur le reposoir où se situe le stimulateur. Il explore ensuite l'environnement avec sa main gauche pour localiser cette cible.

| Groupe | Cibles | Entraînement préalable |
|---|---|---|
| 10 personnes aveugles de naissance | DIRECTION | NON |
| 10 sujets voyants aveuglés | DIRECTION | NON |
| 10 sujets voyants aveuglés | DISTANCE | NON |
| 10 sujets voyants aveuglés | DIRECTION | OUI |
| 10 sujets voyants aveuglés | DISTANCE | OUI |

*Tableau 1: Constitution des groupes de sujets ayant participé à l'expérience.*

Au bout de ces dix essais effectués de façon autonome par le sujet, l'expérimentateur place la cible sur la table, demande au sujet de l'explorer puis substitue à la cible un disque en papier d'un diamètre large d'un centimètre de plus que celui de la cible. Le sujet doit pointer sur la position estimée de la cible. La phase d'entraînement prend fin lorsque le sujet est capable de pointer dix fois à sur le disque en papier avec son index droit. Si ce n'est pas le cas la procédure est répétée avant que l'expérience en elle-même ne commence. La durée de cette phase d'entraînement n'a jamais dépassé 20 minutes. Soulignons que durant cette phase d'entraînement et pendant elle seule, les sujets ont l'occasion de mettre en correspondance de façon active une localisation proprioceptivement connue de la cible avec une exploration consécutive à l'aide de la prothèse[5]. Les sujets de ces groupes et eux-seuls ont l'occasion d'observer les conséquences sensorielles (via la prothèse) des déplacements de la cible sur le point de travail. Ils ont donc un retour sur les conséquences perceptives des modification qu'ils induisent dans l'environnement.

A chaque essai, l'expérimentateur relève donc la réponse verbale (GO ou le chiffre correspondant à la position de la cible) ainsi que la position de l'index de la main droite à la fin du mouvement de pointage (avec une précision de un centimètre). La précision du pointage est mesurée par la distance entre la position de la cible et la position relevée par l'expérimentateur, ou par la différence entre la direction/distance du pointage par rapport à celle de la cible.

## Résultats

*Quelle est la précision de la localisation ou de la discrimination des positions des cibles ?*

Cette précision est relativement bonne puisqu'elle est de l'ordre du diamètre de la balle de tennis. Elle dépend de la position de cible considérée et de la disposition globale des cibles. Cette précision est très altérée pour les cibles les plus éloignées pour les groupes DISTANCE. Ce résultat indique que la localisation distale d'une cible avec le dispositif n'est bonne (sans apprentissage préalable) que dans l'espace proximal du sujet.

*L'introduction d'une période préalable d'apprentissage améliore-t-elle la précision des sujets ?*

La réponse est sans équivoque oui. La précision est grandement améliorée en particulier dans le cas des groupes DISTANCE pour les cibles les plus éloignées. Si l'on effectue une analyse statistique (ANOVA) de la distance moyenne à la cible en considérant les deux facteurs CIBLE (6 niveaux) et ENTRAINEMENT (2 niveaux), on obtient dans le cas de l'expérience « direction », un effet significatif du facteur ENTRAINEMENT ($F[1,18] = 9,17$; $p<0,01$), dans le cas de l'expérience « distance », un effet significatif du facteur CIBLE ($F[5,90] = 10,94$; $p<0,00001$) et un effet significatif du facteur d'interaction

---

5. Notons également que si la tâche verbale (condition CHIFFRE) est clairement une tâche de discrimination de position de cible, la tâche de pointage peut être considérée soit également comme une tâche de discrimination (le sujet répond en pointant sur la position proprioceptivement mémorisée qu'il associe à la position de la cible présentée) soit comme une tâche classique de pointage (le sujet pointe là où il perçoit que la cible se trouve).

CIBLExENTRAINEMENT (F[5,90]=2,42; p<0,05). Ceci indique que la courte période d'apprentissage améliore la précision quelque soit la position de la cible dans l'expérience « DIRECTION » alors que dans l'expérience « DISTANCE », c'est la précision du pointage pour les cibles les plus éloignées qui est améliorée.

### *Quelles sont les caractéristiques du référentiel dans lequel la position des cibles est codée ?*

Déterminer les caractéristiques du ou des référentiels hypothétiques dans lesquelles les position des cibles seraient codées est une question récurrente dans la littérature concernant l'étude des tâches de pointage. Il est admis que la distribution des lieux de pointage dépend à la fois du mode de présentation des cibles (mode visuel ou proprioceptif), de l'introduction d'un délai entre la présentation de la cible et la réponse du sujet et de l'association ou non au pointage d'une verbalisation indiquant le numéro de la cible pointée. Pour une cible donnée, il est classique de mesurer pour chaque sujet les erreurs constantes et variables[6]. commises et les caractéristiques de l'ellipse de confiance associée à la cible considérée. Pour des cibles disposées selon un arc de cercle (comme dans l'expérience DIRECTION) les ellipses associées à des cibles présentées visuellement auront leur grand axe plutôt aligné avec la direction de la cible. Pour des cibles présentées proprioceptivement[7], le grand axe sera approximativement orthogonal à celui du cas précédent, c'est à dire orthogonal à la direction dans laquelle se situe la cible (avec beaucoup de variabilité dans l'orientation de cet axe).

C'est exactement l'inverse pour des cibles alignées comme dans l'expérience DISTANCE. L'argument avancé par certains auteurs pour expliquer ce phénomène est que le système visuel (en vision monoculaire) serait relativement mauvais dans l'estimation des distances et plutôt performant dans l'estimation des directions, et que c'est l'inverse pour le membre supérieur, du fait de la géométrie du bras [BEERS 1998][BEERS 1999]. Si l'on introduit un délai de l'ordre de quelques secondes entre la présentation proprioceptive d'une cible et la réponse pointée du sujet, on observe que la précision du pointage est moins bonne, mais surtour que l'ellipse de confiance voit son grand axe adopter une orientation systématiquement orthogonale à la direction de la cible.

Pour un ensemble de cibles disposées radialement, les ellipses de confiance prennent alors une disposition régulière en arc de cercle. On obtient le même type de disposition si l'on demande au sujet de pointer immédiatement après la présentation proprioceptive de la cible mais en lui demandant de verbaliser un chiffre correspondant au numéro de la position. L'interprétation proposée est la suivante : lorsque le délai imposé dépasse la capacité d'une mémoire motrice limitée temporellement, celle-ci est « contaminée » par des informations provenant de processus cognitifs de plus haut niveau. La position pointée par le sujet n'est plus seulement une production de la mémoire motrice, mais tient alors compte de la configuration de l'environnement (le contexte). Dans ce cas, les positions pointées par les sujets sont influencées par les disposition relatives des cibles [ROSS 1999][ROSS 1996]. La réponse du sujet s'appuierait alors sur une mémoire spatiale utilisant un référentiel allocentré. Si l'on demande au sujet de verbaliser en même temps qu'il pointe le numéro de la cible, alors l'utilisation d'un référentiel allocentré serait forcée par la contrainte pour le sujet d'élaborer une une représentation « sémantique » de la position de la cible [PIS 1996].

---

6. L'erreur constante est la différence entre la position moyenne pointée par le sujet et la position de la cible. L'erreur variable est l'écart-type des pointages autour de cette position moyenne (écart-type du pointage moyen). L'ellipse de confiance représente la zone elliptique en dehors de laquelle un pointage a statistiquement moins de 5 chances sur cent de se localiser.
7. Selon les études, la cible est présentée « proprioceptivement » soit en demandant au sujet de repointer à l'endroit ou l'expérimentateur a placé son doigt, soit en lui donnant comme consigne de pointer avec l'index d'une main la position pointée sous le plan de travail par l'index de l'autre main.

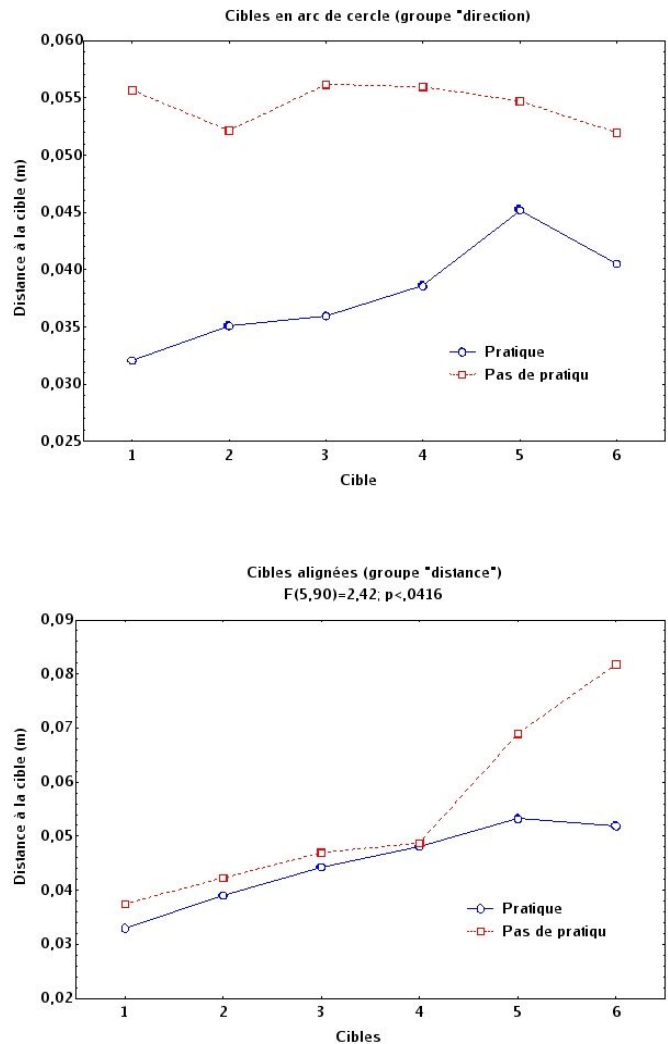

*Figure 3: Comparaison des distances moyennes à la cible pour les groupes ayant eu accès ou non à une préiode de pratique (entraînement) avec le dispositif.*

Qu'en est-il des sujets utilisant le dispositif de Lenay pour percevoir la position des cibles ? Sur la figure 4 sont représentées pour l'expérience DIRECTION les ellipses de confiance pour les deux groupes de sujets (sujets ayant eu droit à une courte période d'entraînement ou non). Ces ellipses ont été calculées en groupant tous les pointages et correspondent donc à la distribution obtenue pour un sujet « moyen ». Il est d'abord clair que la présence de la période d'entraînement diminue simultanément la longueur des grands et petits axes des ellipses et donc leur surface. Mais on observe également que les grands axes des ellipses sont approximativement orthogonaux à la direction de la cible. Si l'on se réfère à la littérature précédemment citée, il apparaîtrait donc que l'action motrice des sujets est (1) guidée par une mémoire spatiale de type allocentré ou/et (2) équivalente à l'action obtenue dans le cas d'une présentation proprioceptive des cibles. Il est difficile dans le cas de cette expérience de conclure sur l'origine de ce phénomène car plusieurs interprétations peuvent être proposées. L'explication la plus simple est que l'exploration de la scène avec le dispositif de Lenay placé sur l'index de la main gauche fait appel à un schème de pointage pouvant se faire très près de la cible. Comme la main droite du sujet repose à plat sur le plan de travail, celui-ci peut déduire de l'intersection entre sa perception de l'orientation du plan de travail et la direction pointée une localisation « proprioceptive » de la cible. On peut également objecter que dans le protocole présenté un délai de réponse de l'ordre de quelques secondes (1-2 s) existe entre la fin de l'exploration du sujet et le signal donné par l'expérimentateur. Ce délai, comme dans le cas des expériences citées précédemment forcerait l'utilisation d'une mémoire spatiale allocentrée.

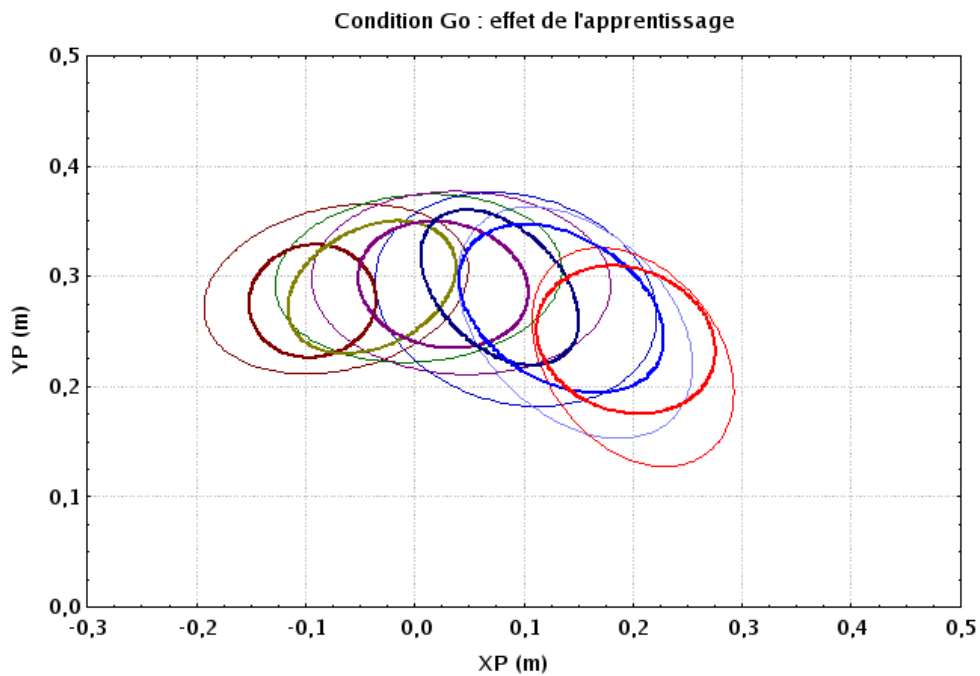

*Figure 4: Ellipses de confiance pour la condition **GO** pour deux groupes ayant passé l'expérience "direction". Seul un des deux groupe (ellipses en gras) a bénéficié d'une phase d'entraînement avec le dispositif avant de passer l'expérience.*

Il est d'autre part intéressant de remarquer une diminution de l'excentricité des ellipses associée à la présence de la période d'entraînement : les ellipses apparaissent plus « rondes » et c'est en particulier vrai pour les ellipses situées en face du sujet. Même si ce changement reste à objectiver c'est peut-être le signe d'une modification induite par l'entrainement sur les processus guidant la perception de la localisation des cibles. C'est sans doute le signe d'une meilleur appropriation du dispositif, les sujets étant capables de l'utiliser de façon à percevoir correctement à la fois la distance et la direction dans laquelle la cible se situe. La cible étant mieux individualisée de par ses caractéristiques spatiales, l'utilisation d'une mémoire allocentrée dépendante du contexte ne se justifie plus.

# Perception proximale de l'orientation d'un objet à l'aide d'un dispositif « virtuel »

## *Objectifs de l'expérience*

Les objectifs de cette expérience préliminaire sont multiples [TALEB 2002]. Il s'agit d'abord de réaliser techniquement et de tester une adaptation de l'utilisation du dispositif de Lenay en environnement virtuel. Un tel environnement permet d'une part de s'affranchir des perturbations présentes dans un environnement réel (ombres, réflections etc...) et d'autre part d'enregistrer les mouvements exploratoires produits par les sujets. Nous cherchons également à examiner la capacité de sujets normaux aveuglés à reproduire l'orientation d'une forme bidimensionnelle déposée sur unplan. Dans cette expérience, la forme est en fait tridimensionnelle (un cylindre) mais son intersection avec la surface explorée par les sujets (un plan de travail) est un rectangle. Des travaux préliminaires (utilisant un autre dispositif) ont montré que des sujets normaux aveuglés étaient capables de reproduire correctement l'orientation de segments de lignes [LENAY 2002]. Mais ces segments étaient dotés d'une épaisseur très faible. Dans cette expérience, le rapport largeur sur longueur de l'objet est bien plus important[8]. Nous souhaitons donc savoir si cette capacité est conservée lorsque les objets possèdent une épaisseur notable. Enfin, si le cylindre est présenté toujours au même endroit sur le plan de travail, nous demandons au sujet de reproduire son orientation en trois endroits différents : près de la position centrale où est présenté l'objet, ou bien à gauche ou à droite de l'axe de symétrie du corps du sujet (voir figure 5). Nous souhaitons en effet tester l'hypothèse selon laquelle le déplacement nécessaire pour atteindre les lieux de reproduction latéraux introduirait une perturbation dans la perception qu'on les sujets de l'orientation du cylindre ou dans la reproduction qu'ils en font.

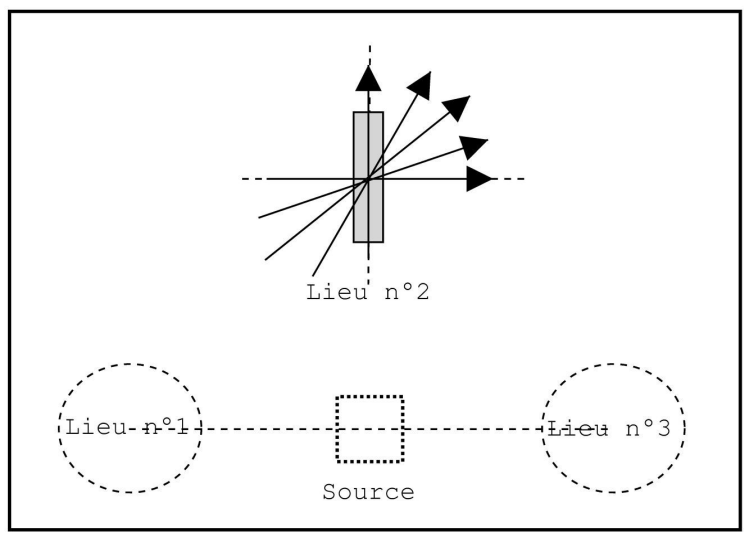

*Figure 5: Ce schéma figure les différentes orientations dans lesquelles sont présentées l'objet (0°, 30°, 45°, 60° et 90°). La source électromagnétique du dispositif de capture du mouvement est fixée sous la table (rectangle en pointillé). Les deux ellipses en pointillés représentent les zones latérales ou l'on demande au sujet de reproduire par un tracé rectiligne l'orientation perçue de l'objet. Les trois lieux sont situés à une distance de 20 cm de la source.*

---

8. Ce rapport est d'un quart : le cylindre fait 20 cm de long pour 5 cm de large)

## *Matériel et Méthodes*

### *Le dispositif de Lenay en environnement virtuel*

Le dispositif expérimental utilise un système de capture des mouvements FastTrack de la société Polhemus avec un unique capteur. La position et l'orientation du capteur sont échantillonnées à une fréquence de 120 Hertz. Lorsque le capteur entre dans le volume du cylindre (cylindre virtuel défini uniquement par ses propriétés géométriques), une carte relais déclenche la mise en mouvement d'un moteur vibreur. Ce vibreur est conditionné dans un quadrilatère en bois plein. Ce système extrêmement simple permet ainsi de définir des objets tactiles virtuels.

### *Déroulement de l'expérience*

Le sujet est assis devant la table, il est aveuglé par un masque et porte des bouchons d'oreille empêchant l'audition des vibrations du stimulateur. Il explore ensuite la surface de la table en tenant le capteur entre le pouce et l'index. La période d'exploration est limitée à 30 secondes. Lorsque cette période expire, l'expérimentateur tape sur l'épaule du sujet qui immobilise alors sa position. Le bras du sujet est alors déplacé passivement vers la zone de reproduction de l'orientation. Au bout de cinq secondes un nouveau signal est fourni au sujet par l'expérimentateur. Le sujet trace alors un segment de droite en maintenant le bouton d'une souris d'ordinateur appuyé. Ces appuis, enregistrés avec les données de capture du mouvement, permet de repérer aisément les phases où le sujet trace ce segment de droite. Le sujet replace activement son bras à la position de départ pour enchaîner ensuite l'essai suivant. L'expérience comprend cinq blocs randomisés présentant toutes les orientations trois fois chacune (une fois par lieu). Un bloc comprend donc 15 essais et chaque sujet (ils sont au nombre de cinq) effectue 75 essais.

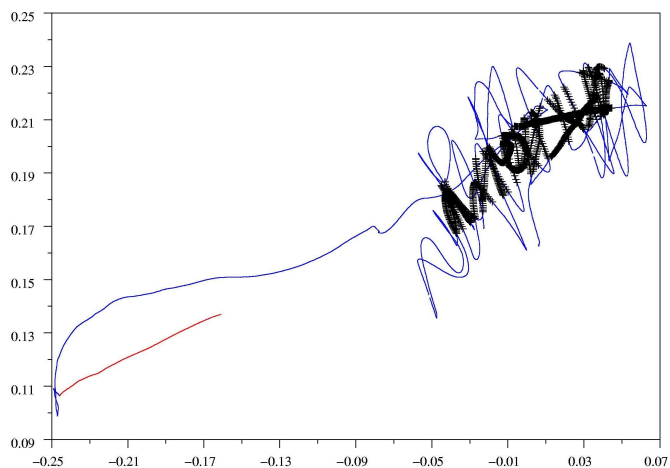

*Figure 6: Graphique reproduisant le déroulement d'un essai. Le fin tracé bleu représente la trajectoire du capteur sur le plan de travail. Les points de contact avec la forme sont illustrés par des points noirs épais. Le tracé effectué par le sujet pour reproduire l'ordintation perçue de la forme est en rouge (lieu n°1).*

Les données numériques recueillies sont donc des séries temporelles comportant à chaque pas de temps la position et l'orientation du capteur ainsi que deux signaux binaires indiquant la présence ou non d'une stimulation tactile et si le sujet est en train ou non de produire son estimation de l'orientation du cylindre (appui sur le bouton de la souris). Un calcul de régression linéaire permet d'obtenir l'angle du segment tracé à chaque essai. Cet angle est comparé à la véritable orientation du cylindre. Nous avons également calculé (toujours par une régression linéaire) l'orientation du nuage des points de contact du capteur avec l'objet. Ce calcul permet d'obtenir l'orientation de l'objet véritablement exploré par le sujet : si celui-ci n'explore qu'une partie du cylindre, il peut parfaitement reproduire l'orientation de cette partie bien qu'elle soit différente de l'orientation globale du cylindre lui-même. On peut donc en s'intéressant à cette orientation du nuage des points de contact, faire la différence entre les erreurs dues à une mauvaise perception de l'objet et celles dues à un mauvais fonctionnement de la mémorisation et de la reproduction de l'orientation de celui-ci.

### *Résultats*

L'erreur RMS angulaire des sujets ne semble pas dépendre pas du lieu de reproduction de l'orientation de l'objet (voir table 2). L'erreur (RMS) entre l'orientation de l'objet et l'orientation proposée par les sujets est très importante, de l'ordre de d'une vingtaine de degrés. Les sujets présentent donc a priori des performances pauvres dans cette tâche. Cette erreur est en tout cas bien supérieure à celle observée dans une expérience similaire [LENAY 2002] impliquant la perception de l'orientation de segments de lignes fines. Le fait que l'objet perçu soit doté d'une épaisseur non négligeable augmente visiblement la difficulté de la tâche. Mais cette baisse de performance est-elle due (1) aux difficultés des sujets à percevoir l'orientation globale de l'objet (insuffisance de la prothèse ou du processus de liage des zones explorées) ou bien (2) à leur incapacité à déterminer l'orientation globale des points de contact avec la forme (insuffisance de l'empant ou de la capacité temporelle de la mémoire sensorimotrice).

| Sujet | Lieu n°1 | Lieu N°2 | Lieu N°3 |
|---|---|---|---|
| 1 | 18.46 | 9.75 | 17.01 |
| 2 | 26.16 | 24.75 | 22.36 |
| 3 | 17.18 | 21.07 | 14.72 |
| 4 | 20.67 | 14.17 | 20.45 |
| 5 | 22.71 | 18.50 | 18.96 |

*Tableau 2: Erreur RMS en degrés dans la reproduction de l'orientation des objets pour les trois lieux de reproduction considérés*

On peut en partie tenter de répondre à cette question en mesurant l'erreur (par exemple erreur RMS angulaire) entre l'orientation de l'objet et l'orientation du nuage des points de contact. Si cette seconde erreur est inférieur à la première, on ne peut mettre en cause la capacité spatio-temporelle de la mémoire ou des mémoires en jeu. Il est alors nécessaire de considérer que les erreurs importantes commises par les sujets ont une autre origine. La nature de la prothèse et le type de couplage sensation-action quelle impose au sujet pourrait bien être impliquée. Nous présentons un exemple dans lequel la réponse du sujet semble ainsi clairement influencée par les mouvements d'exploration qu'il a effectué pour percevoir la forme cylindrique présentée (figure ). Il serait donc intéressant de comparer la force de la corrélation entre l'orientation du cylindre et les réponses du sujet à la force de la corrélation entre l'orientation des points de contact (ou des mouvements de balayage) et les réponses du sujet.

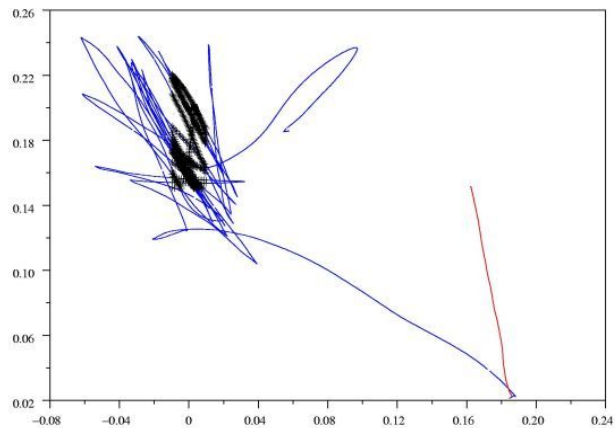

*Figure 7: Exemple d'un essai ou le cylindre présente une orientation de 90 degrés, mais ou la réponse du sujet semble influencée par son anticipation (orientation des balayages) sur la forme.*

On retrouve la structure du couplage sensori-moteur impliqué dans le microbalayage déjà observée dans le cas d'un dispositif se servant d'une tablette graphique pour enregistrer les mouvements des sujets [HANN 1998]. Les points de contact avec la forme sont ainsi distribués au niveau des pics de la vitesse tangentielle du mouvement (voir figure  ). La vitesse tangentielle des pics est relativement faible (de l'ordre de 20 centimètres par seconde) en tout cas bien inférieure à la vitesse maximale de balayage que peut produire le bras humain. La fréquence fondamentale des balayages est également faible (environ 1 Hertz). Enfin, lorsque le sujet semble anticiper son parcours sur la forme, la vitesse tangentielle moyenne augmente régulièrement. Tout se passe comme si le sujet introduisait une accélération dans son activité exploratoire. La vitesse de l'exploration est donc contrainte à la fois par le couplage induit par le dispositif et par les mécanismes anticipatoires en jeu. Le sujet règle sans doute sa vitesse pour « écouter » de façon optimale le dispositif et celle-ci est sans doute également influencée par les processus de nature plus cognitifs en jeu dans le liage des différentes parties de la forme explorée.

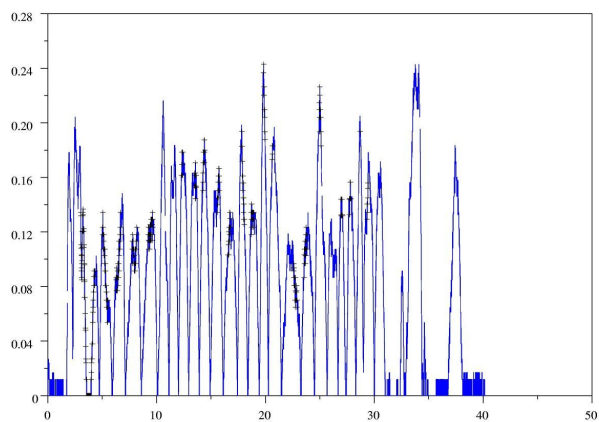

*Figure 8: Tracé de la vitesse tangentielle (mètres par seconde) en fonction du temps (secondes) pour l'essai présenté en figure 7. Les vitesses tangentielles aux points de contact sont marquées par des croix.*

Les résultats obtenus avec cette expérience valident, par leur similitude avec des résultats précédemment établis, l'utilisation d'un environnement virtuel pour la simulation de dispositifs de Lenay. Ce type d'environnement autorise la simulation de prothèses proximales et distales en trois dimensions et présente l'avantage de donner un accès aux variables cinématiques décrivant les trajectoires perceptives produites par les sujets. Cette approche permet ainsi de donner accès aux détails cinématiques des couplages sensori-moteurs induits par les différents types de prothèses. Mais encore faut-il se donner des outils méthodologiques adéquats pour analyser les données recueillies.

# Conclusion

Nous espérons avoir convaincu le lecteur de l'intérêt du dispositif pour explorer les mécanismes intimes de l'activité perceptive dans le cadre de la suppléance sensorielle. Nous espérons également avoir montré l'intérêt de la transposition de méthodes issues de la psychologie expérimentale à l'étude de la perception techniquement médiée[9]. L'intérêt majeur d'utiliser la forme simple de la prothèse est celui de mettre le sujet dans l'obligation d'articuler finement sensori-motricité et processus de nature plus cognitive. Si l'on s'en tient à la littérature sur le sujet, dans la première expérience présentée la distribution des pointages des sujets reflète certainement l'interaction entre mémoire sensori-motrice et structure de l'imagerie mentale. Selon une proposition récente, la nature même de la sensation perçue par les sujets (sonore, visuelle etc...) utilisant un système de suppléance perceptive serait dictée par les contingences sensori-motrices induites par le port du dispositif [OREGAN 2001]. Avec le dispositif utilisé, la sensation de présence de l'objet peut être ressentie par les sujets comme visuelle (le phototransistor réagit à la lumière), tactile (les stimulations sont tactiles), ni l'un ni l'autre ou entre les deux. Mais comment objectiver le fait qu'un sujet utilisant un dispositif de suppléance particulier perçoit un objet comme s'il le touchait, l'entendait ou le voyait ? L'analyse de la forme des distributions des pointages peut peut-être permettre de répondre à cette question puisque cette distribution semble spécifique de la modalité de présentation de la cible. Nous proposons également que la forme de la distribution reflète le degré d'intégration du dispositif.

Les caractéristiques de la mémoire sensori-motrice de travail apparaissent comme centrales dans ce contexte. En effet, les capacités de perception du sujet dépendent crucialement de son aptitude à lier une séquence d'actions exploratoires à une séquence de sensations élémentaires. L'analyse des trajectoires exploratoires et en particulier de la structure des micro-balayages pourrait s'avérer très informative de ce point de vue. Le micro-balayage, qui est incessant tant que le sujet ne perd pas le contact avec la forme, a sans doute entre autre pour rôle de « rafraîchir » en permanence cette mémoire afin de la maintenir accessible au processus de liage. La fréquence d'oscillation des mouvements (un élément de micro-balayage dure environ une à deux secondes) peut être mis en rapport avec le délai nécessaire dans les expériences de pointage pour que la mémoire motrice du sujet soit perfusée par des processus cognitifs de plus haut niveau [ROSS 1999]. Si cette fréquence d'oscillation reste à peu près constante, il n'en est pas de même pour la vitesse du mouvement. Celle-ci peut être décomposée en deux signaux : un signal haute fréquence correspondant aux micro-balayages superposé à un signal de plus basse fréquence lié au suivi des contours de la forme explorée (de direction orthogonale à celle du premier). Si l'on accepte ce modèle, des changements au niveau de ce signal basse fréquence pourraient être responsables de l'augmentation de la vitesse tangentielle moyenne observée lorsque le sujet anticipe correctement les contours de la forme explorée. La présence d'une accélération dans ce signal serait ainsi le signe que le sujet anticipe correctement les conséquences sensorielles de ses actions : *il sait comment et où il peut sentir quoi*. Il perçoit.

---

9. Même si les concepts manipulés par les psychologues paraissent parfois à certains réducteurs ou bien inadaptés (notion de référentiel et de représentation interne) les méthodes utilisées en sont relativement indépendantes.

# Références